\documentstyle[aps,multicol,epsf,epsfig]{revtex}

\begin{document}

\title{Free cooling and inelastic collapse of granular gases in high dimensions}

\author{Emmanuel Trizac\footnote{Electronic Address: 
Emmanuel.Trizac@th.u-psud.fr}
and Alain Barrat\footnote{Electronic Address: Alain.Barrat@th.u-psud.fr} }

\address{
Laboratoire de Physique 
Th\'eorique\footnote{Unit\'e Mixte de Recherche UMR 8627 du CNRS}, 
B\^atiment 210\\ Universit\'e de Paris-Sud, 91405 Orsay Cedex, France
}
 
\date{\today}

\maketitle 
\begin{abstract}
The connection between granular gases and sticky gases has recently been
considered, leading to the conjecture that inelastic collapse is avoided 
for space
dimensions higher than 4. We report Molecular Dynamics simulations of
hard inelastic spheres in dimensions 4, 5 and 6. The evolution of the 
granular medium is monitored throughout the cooling process. The behaviour 
is found to be very similar to that of a two-dimensional system, with
a shearing-like instability of the velocity field and inelastic collapse
when collisions are inelastic enough, showing that the connection
with sticky gases needs to be revised. 
\end{abstract}

\vskip 5mm


An important difference between a molecular fluid and a gas of mesoscopic or
macroscopic grains is the possibility for the latter, associated 
with the inelastic nature of the
collisions, to exhibit clustering and collapse 
\cite{Goldhirsch,McNamara,Luding}.
A large and rapidly growing body of theoretical work is devoted to
clustering, which consists in a long wavelength low frequency hydrodynamic
phenomenon, and refers to the formation of density inhomogeneities.
On the other hand, the phenomenon of inelastic collapse, which is 
a short wavelength and high frequency singularity inherent to the 
inelastic hard sphere (IHS) model, seems much less understood,
except in one dimension \cite{Bernu}.

In the IHS model, grains are modeled as smooth hard spheres undergoing
binary, inelastic and momentum-conserving collisions,
which dissipate a constant fraction $(1-r)$ of the component of
the relative velocity ${\mathbf v}_{12}$ along the
center-to-center direction $\bbox{\hat \sigma}$. 
Noting with primes the post-collision velocities, this translates
into ${\mathbf v}_{12}' \cdot \bbox{\hat{\sigma}}
= -r \,{\mathbf v}_{12} \cdot \bbox{\hat{\sigma}}$, while 
the tangential relative
velocity (perpendicular to $\bbox{\hat\sigma}$) is conserved.

\begin{figure}[ht]
\centerline{ 
\epsfig{figure=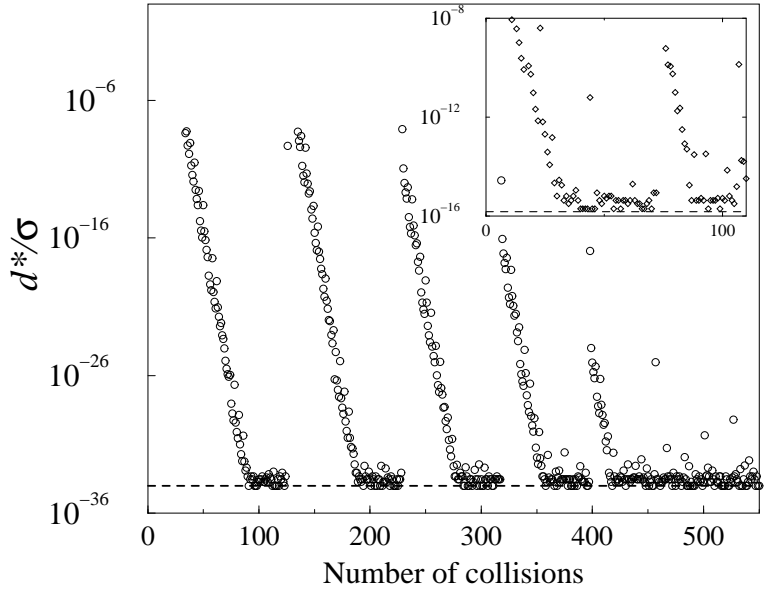,width=10cm,angle=0}    }
\vskip 2mm       
\caption{Normalized inter-particle separation (see text for definition) 
as a function of the number of collisions since an arbitrary time origin, 
for $d=5$, $N=16807$, $\phi=0.08$ and $r=0.1$. Each circle corresponds to 
a collision and the data were obtained specifying a ``quadruple precision''
computation (with reals coded on 16 bytes). 
Each decrease (from $10^{-8}$ to $10^{-36}$) corresponds to repeated collisions
between a small number of particles, typically three particles, one bouncing
back and forth between two others, with a diverging frequency.
For the same system, the inset 
shows with diamonds the results of a standard ``double precision'' 
run (reals on 8 bytes). In both cases, the floor of machine precision
is indicated by a dashed line (approximately $10^{-34}$ for real$\star$16 and
$10^{-16}$ for real$\star$8). Only those collisions with $d^*<10^{-8} \sigma$ 
have been displayed.}
\label{fig:collapse}
\end{figure}

In an interesting article, Ben-Naim {\it et al.} proposed that a 
freely evolving inelastic gas belongs asymptotically to the universality
class of the sticky gas \cite{BenNaim},
for which ${\mathbf v}_{12}'=\bbox{0}$ after each collision. Noticing that 
the temperature $T$ of an inelastic gas is a monotonically increasing function 
of the restitution coefficient $r$ and therefore bounded from below by the
totally inelastic case ($r=0$), these authors invoked a mapping onto
Burgers' equation
to conjecture that the inelastic collapse is avoided for space dimensions
$d>d_c = 4$, and that the standard Haff's cooling law 
$T \propto (\epsilon t)^{-2}$
where $\epsilon=(1-r^2)/(2d)$, holds indefinitely above this critical
dimension $d_c$.
Velocity fluctuations and scaling exponents in one dimension 
can indeed be described by the inviscid Burgers
equation \cite{BenNaim}. However, in higher dimensions, 
the completely inelastic version of the
IHS model, with $r=0$, does not strictly correspond to the
sticky gas limit because the tangential relative velocity is not dissipated 
in a binary encounter (${\mathbf v}_{12}' \cdot \bbox{\hat{\sigma}}=
{\mathbf 0}$, but a priori ${\mathbf v}_{12}' \ne {\mathbf 0}$),
so that it is interesting to test the validity 
of the above-mentioned predictions. In this article, we report 
Molecular Dynamics (MD) simulations of IHS gases 
for $d=4,5$ and 6. For each space dimension, the relevant parameters
$\phi$ (packing fraction) and $r$ (normal restitution)
are varied for systems consisting typically of $N=5.10^3$ to $5.10^4$ 
particles. 
The code was first successfully tested by comparing for various densities
the MD equation of state (or equivalently the pair correlation function 
at contact) with the analytical approximation of Song {\it et al}
\cite{Song}.
For all investigated dimensions and for high enough dissipation
($r \leq 0.2$ for $d=5$ and $r \leq 0.1$ for $d=6$), the system exhibits
the finite time singularity characteristic of the inelastic collapse,
in contradistinction to the conjecture of \cite{BenNaim},
with a situation closely reminiscent to its two dimensional 
counterpart \cite{McNamara}: the (hyper)spheres collide infinitely
often in a finite time along their joint line of centers. Following 
Refs. \cite{McNamara} we probed this multi-particle process occurring
through the accumulation of an infinite sequence of binary collisions
by a contact criterion: after each collision, the relative distance $d^*$
between the 
next two colliding partners is monitored; if this interparticle spacing
normalized by the diameter $\sigma$
has decreased and becomes of the order of machine precision, 
a three body interaction has occurred, corresponding to an
inelastic collapse.
The results of a typical run ($d=5$) are shown in 
Fig. \ref{fig:collapse}. When a multi-body interaction commences, 
$d^*$ decreases geometrically with the number of 
collisions, as in two dimensions \cite{McNamara}.
Enforcing a high precision computation allows to follow the 
decay over more than 26 orders of magnitude (whereas only 
8 orders are accessible with a standard double precision algorithm, 
see the inset of Fig. \ref{fig:collapse}). After a collapse
has occurred, the inaccuracy of the computer disperses the collapsing
cluster, before another multi-body event involving different particles
occurs at a different location. Our analysis indicates that on 
a hypothetical infinite precision machine, the collapse would continue
forever whereas roundoff errors act as an effective regularization.
Throughout a collapse, the time between two successive collisions follows
a geometrical decrease very similar to that displayed in Fig.
\ref{fig:collapse}.

Let us note that the seemingly low value of the packing fraction $\phi$ in Fig.
\ref{fig:collapse} is a misleading effect of ``high'' 
dimensionality. It turns that the reduced density 
\begin{equation}
n^*\, = \,n \,\sigma^d \,= \,\frac{d \, 2^{d-1}}{ \pi^{d/2}}\, 
\Gamma\left(\frac{d}{2}\right) \phi
\end{equation}
where $n$ is the number density and $\Gamma$ the Euler function, 
is a more relevant parameter to discriminate between ``low'' and 
``high'' densities. In a simple cubic lattice, the highest packing
achievable with spheres at contact corresponds to $n^*=1$. The configuration
of Fig. \ref{fig:collapse} with $\phi=0.08$ corresponds to
$n^*=0.5$ and is consequently of high density.
We considered the possibility that the conjecture of \cite{BenNaim}
applies in the opposite limit of low packing (that would correspond to
the so called sticky dust in the case of  vanishing both
normal and tangential restitution coefficients). 
We lowered $\phi$ by an order of magnitude which requires to consider
large systems to avoid a spurious increase of the mean free path;
our simulations with $N=2.10^5$ particles and $\phi=0.008$ ($n^*=0.05$)
in five dimensions
nevertheless exhibit inelastic collapse for $r<0.05$.

\begin{figure}[ht]
\centerline{ 
\epsfig{figure=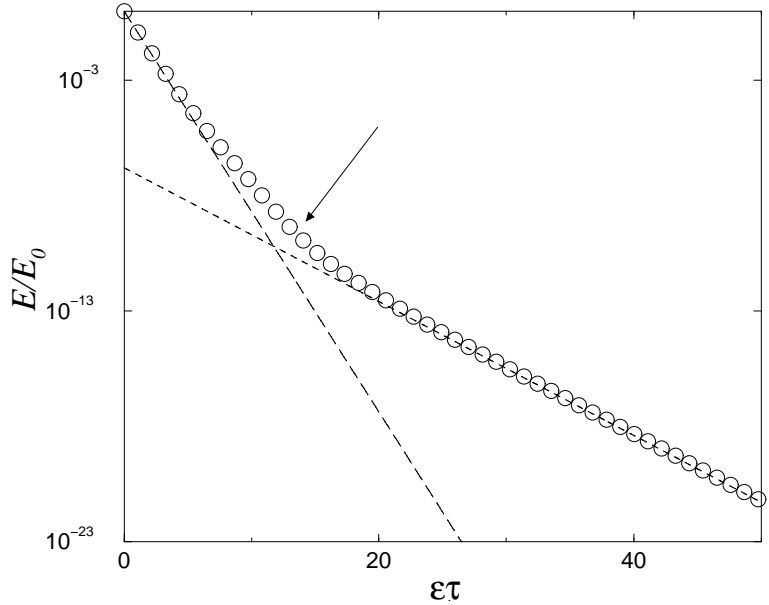,width=10cm,angle=0}    }
\vskip 2mm       
\caption{Evolution on a linear-log scale 
of the mean kinetic energy (rescaled by its initial
value) with the number of collisions per particle for
$N=16807$, $r=0.6$ ($\epsilon=0.064$),
$\phi=0.08$ and $d=5$. The long-dashed line
has a slope $-2$ (Haff's law) while the short-dashed one is a fit
to the long time behaviour with slope $-0.66$.
}
\label{fig:crossover}
\end{figure}

\begin{figure}[ht]
\centerline{ 
\epsfig{figure=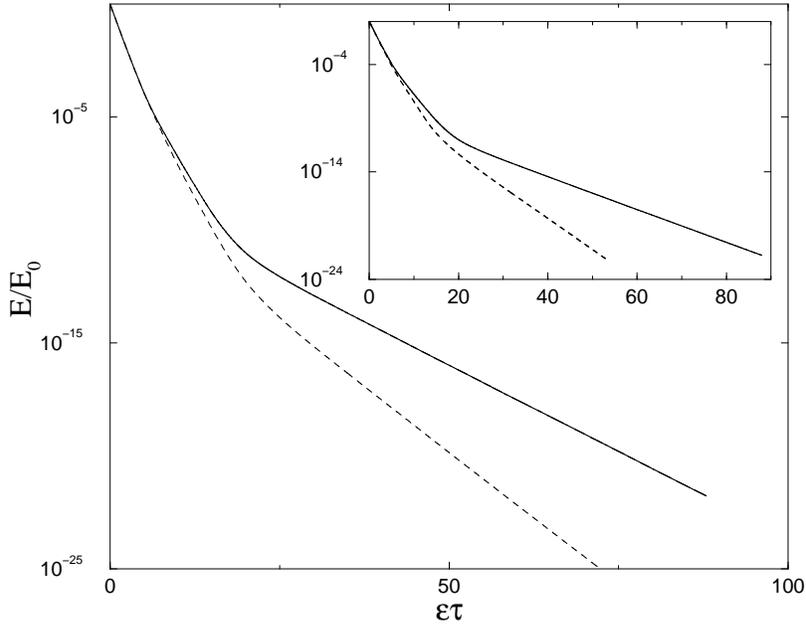,width=8.5cm,angle=-90}    }
\caption{Linear-log plot of energy versus number of collisions for 
$r=0.4$ ($\epsilon=0.084$), 
$\phi=0.08$, $d=5$ and two system sizes : 
$N=16807$ (full curve) and $N=7776$ (dashed line). 
The inset shows the dependence with inelasticity for $N=16807$ :
$r=0.4$ ($\epsilon=0.084$), full curve, and $r=0.6$ ($\epsilon=0.064$),
dashed line. The energy decay thus depends on the system size 
and is not only a function of $\epsilon\tau$, as suggested in 
\protect \cite{BenNaim}.
}
\label{fig:etau}
\end{figure}

\begin{figure}[ht]
\centerline{ 
\epsfig{figure=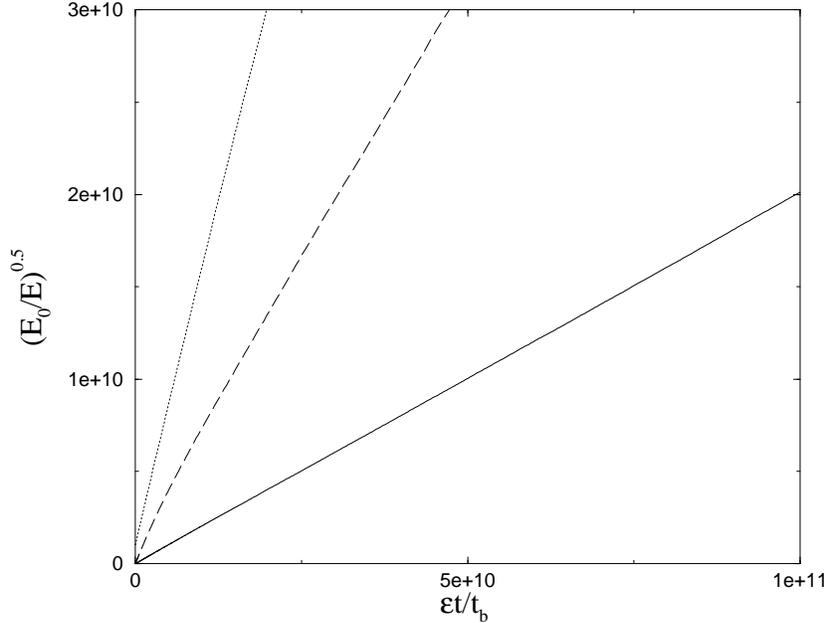,width=8.5cm,angle=-90}    }
\caption{Time dependence of $\sqrt{E_0/E}$ for $r=0.5$
($\epsilon=0.075$) and $d=5$.
The full and long-dashed curves correspond to the large time behaviour 
for $N=16807$ and $N=7776$ respectively. The dotted line with slope 
1.7 is a magnification of the short time evolution, independent of
$N$ and $\epsilon$ (since the evolution follows Haff's law) 
obtained by rescaling 
both $x$ and $y$ axis by the same (large) factor. In the $x$ axis, 
$t$ is expressed in units of $t_b/\epsilon$ where $t_b$ is the Boltzmann
mean collision time of elastic particles at the same 
packing fraction ($\phi=0.08$). These curves show that, although 
$\sqrt{E_0/E}$ increases like $\epsilon t/t_b$, the 
prefactors at short and large times differ, 
and depend at large times on $N$ and $\epsilon$. 
}
\label{fig:et}
\end{figure}

When the restitution coefficient is above a critical threshold,
the inelastic singularity is absent and we can follow the time
evolution of the system.
Initial conditions for all inelastic runs were equilibrated fluid
configurations of elastic hard spheres at the corresponding (uniform) density,
with a uniform
temperature (coarse-grained kinetic energy) and Gaussian distribution of 
velocities. The short time regime is then given by the homogeneous 
cooling state, which is essentially an adiabatically changing
equilibrium state where the mean kinetic energy $E$ (related to the temperature
$T$ by $E=d T/2$ due to the vanishing of the flow field) is given by
\begin{equation}
E(t) \,=\, \frac{E_0}{\left(1+\epsilon t/t_0\right)^2}.
\label{eq:Edet}
\end{equation}
In Eq. (\ref{eq:Edet}), $t_0$ is the Enskog mean collision time
of elastic particles at the same density \cite{Resibois,Deltour} and the 
inelasticity parameter $\epsilon = (1-r^2)/(2d)$ follows from the assumption
of a Gaussian velocity distribution. It has been shown
within the framework of Enskog-Boltzmann equation that the corrections due
to non Gaussian behaviour on the cooling rate are small for all
inelasticities \cite{Twan}. The internal clock of the system
can be parameterized by $\tau$, the average number of collisions suffered
per particle in a time $t$, which reads in the homogeneous cooling
state
\begin{equation}
\tau = \frac{1}{\epsilon}\,\ln\left(1 + \epsilon \,\frac{t}{t_0} \right).
\label{eq:taudet}
\end{equation}
Denoting $N_c$ the total number of collisions having
occurred in the system over a time
$t$, we have $\tau = 2 N_c/N$.
Figure \ref{fig:crossover} shows that the law 
$E(\tau) = E_0 \exp(-2 \epsilon \tau)$
expected from Eqs. (\ref{eq:Edet}) and (\ref{eq:taudet}) is only valid at short
times. A crossover, indicated in Fig. \ref{fig:crossover} by an arrow,
is generically
observed for all runs in dimensions 5 and 6, provided the inelastic collapse 
is avoided; the crossover time depends both on the
inelasticity and on system sizes. 
After the crossover, the behaviour of $E$ with the number of collisions
also depends on system size (see Fig. \ref{fig:etau}),
and $[\ln(E/E_0)]/\tau$ does not scale
with the inelasticity as $\epsilon$ (inset of Fig. \ref{fig:etau}).
Moreover, Fig. \ref{fig:et} shows that the energy decays like $t^{-2}$,
as predicted in \cite{BenNaim}; however, Fig. \ref{fig:et} shows that 
the prefactor is different at short and large times\footnote{We note that 
from Eq. (\ref{eq:Edet}), the slope of the
dotted line in Fig. \ref{fig:et} (short time behaviour) 
corresponds to the ratio
$t_b/t_0$ of Boltzmann over Enskog mean collision times, and is an indirect 
measure of the pair correlation function at contact $\chi$ \cite{Resibois}.
From Fig. \ref{fig:et}, we get $\chi \simeq 1.7$ while the equation of state
of Ref. \cite{Song} yields 1.74 for the density considered. This value
is corroborated by an independent elastic run where we measure 
$\chi \simeq 1.7$.}, and (at large times)
strongly increases with increasing number of particles; it neither scales with 
inelasticity as $\epsilon^{-2}$ (not shown). All these results 
are at variance with the suggestions of \cite{BenNaim}: the
energy at large times decays as 
$A(N,\epsilon) \epsilon^{-2} t^{-2}$,
and not simply as $\epsilon^{-2}t^{-2}$ (Haff's law).

\begin{figure}[ht]
\centerline{ 
\epsfig{figure=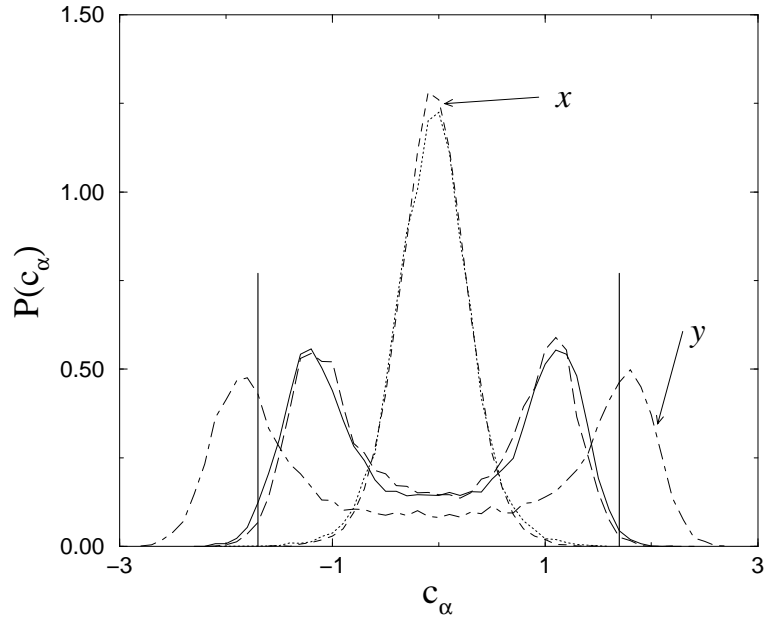,width=10cm,angle=0}    }
\vskip 2mm       
\caption{Probability density distribution of the five components of the
rescaled velocity, for $N=7776$, $r=0.6$, $d=5$ and $\tau=122$. 
One of the central
Gaussian directions is arbitrarily chosen as the $x$ direction and 
that containing
the most significant fraction of the kinetic energy is labeled $y$. The vertical
lines at $c_\alpha = \pm 1.7$ indicate the cutoffs used in the projection scheme
producing Fig \ref{fig:shear}.}
\label{fig:distrvit}
\end{figure}

\begin{center}
\begin{figure}[h]
\centerline{\hspace{-8cm}\psfig{figure=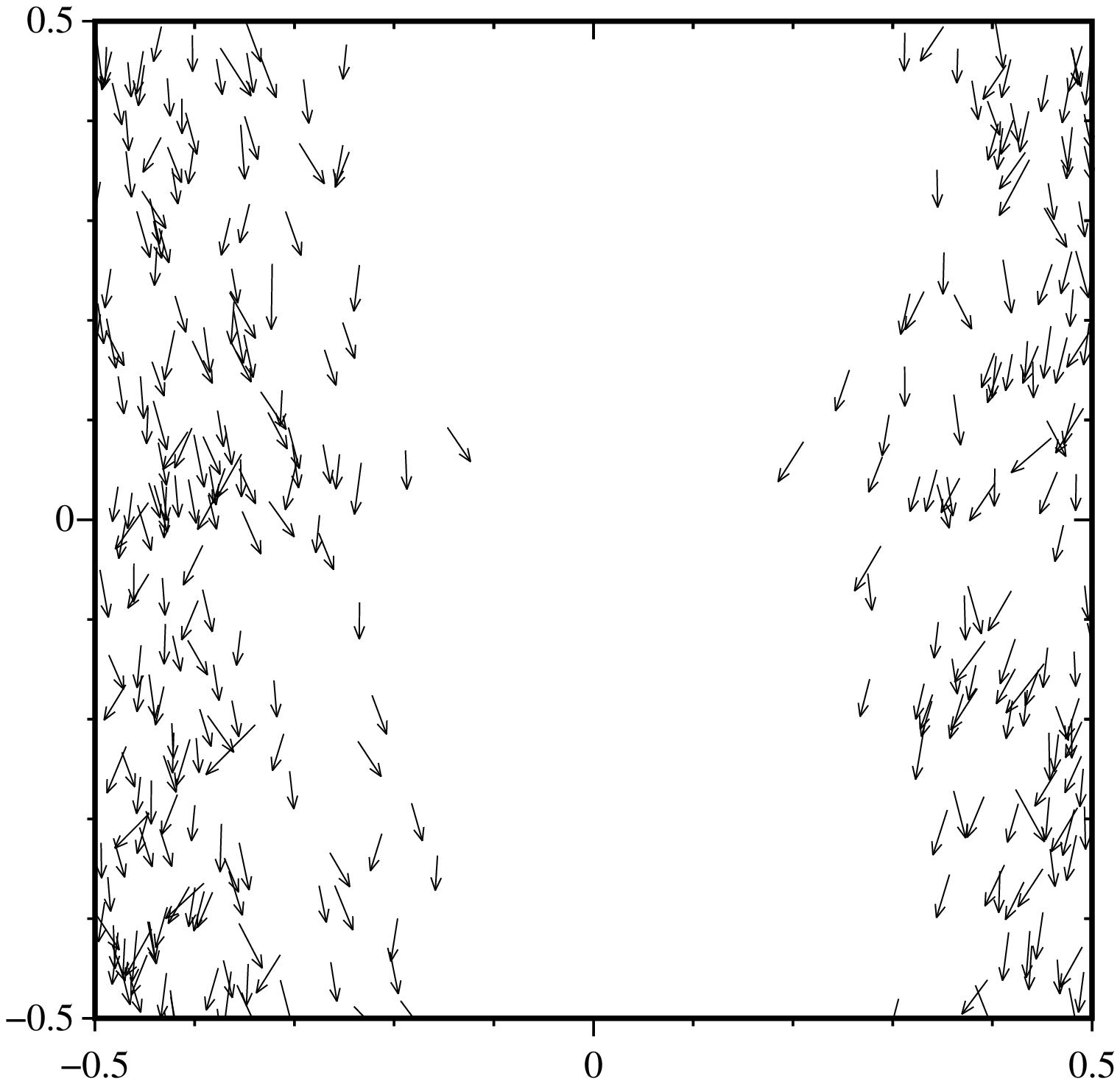,height=13cm}}
\vspace{-13cm}
\centerline{\hspace{8cm}\psfig{figure=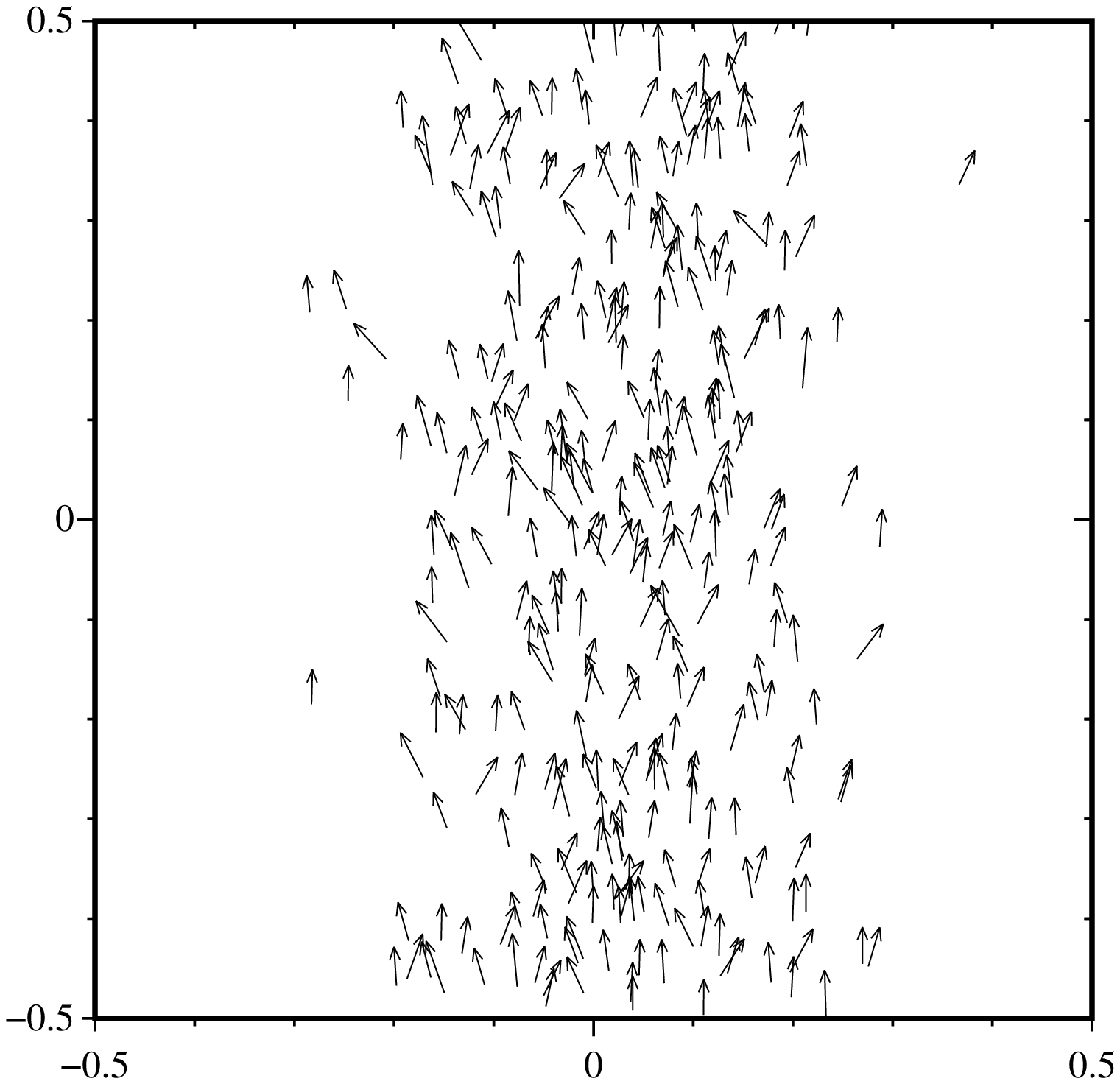,height=13cm}}
\vskip -5mm       
\caption{Projection of the velocities onto the $x$-$y$ plane, for the same system
as in Fig. \ref{fig:distrvit}. Only those particles with $c_y <- 1.7$ (resp.
$c_y>1.7$) have been shown on the left (resp. right) plot. The simulation box is a
cube $[-0.5,0.5]^5$.
\label{fig:shear}}
\end{figure}
\end{center}

Before reaching the crossover time $t_c$, the velocity distribution of the 
particles is isotropic and very close to a Maxwellian. Once the crossover time
is elapsed, we observe an evolution reminiscent of the shearing instability
found in lower dimensions \cite{Goldhirsch,McNamara,Deltour,Orza1}.
Fig. \ref{fig:distrvit} shows the distribution of the components
of the rescaled velocity ${\bf c} = {\bf v}/\sqrt{T(t)}$ for $t \gg t_c$.
For the particular system of Fig. \ref{fig:distrvit},
two components are found to be Gaussian (the central peaks)
and the remaining directions appear as bimodal and concentrate most of the 
energy. The corresponding shear bands are difficult to visualize,
but insight can be gained by a suitable projection 
onto a two-dimensional surface. Such a projection is displayed in Fig. 
\ref{fig:shear}. The coherent motion we observe in Fig. \ref{fig:shear}
suggests that the large time dynamics of the system is controlled
by the periodic boundary conditions used in the simulations, 
as analyzed in \cite{Orza1}.

Since, in a 5 dimensional system of $5.10^4$ particles,
with packing fraction 0.08, the simulation box length $L$ is only 10
times as large as the spheres' diameter, the systems considered here
can be considered as small (in the sense that there exists no
possible separation of length scales). The behaviour reported here for $E(t)$ 
is therefore consistent with the two dimensional 
si\-mu\-lations of Orza {\it et al.} \cite{Orza}, showing 
$E \propto A(N,\epsilon)\, t^{-2}$ for small systems,
with a prefactor much larger than predicted by Haff's law.
Gaining one order of magnitude for $L/\sigma$ would
require to increase $N$ by $d$
orders of magnitude, which is unachievable for $d>4$. The validity of 
Brito and Ernst's theoretical prediction for large systems 
$E \propto \tau^{-d/2}$ \cite{Brito} (coinciding with the approach of
\cite{BenNaim} for $d\leq 4$) can consequently not be tested in dimensions
higher than 4.

In conclusion, our simulations show the existence of an inelastic collapse 
in high dimensions, not only for dense systems but also for more dilute ones 
(even if the situation of very low packing fraction cannot be reached, so that 
the ``sticky dust'' limit cannot be tested by numerical simulations). They 
indicate therefore that the relationship between inelastic and sticky gases 
put forward in \cite{BenNaim} needs to be refined.

It is a pleasure to thank M. Ernst for his continued interest in our work.

\newpage

\vskip -4cm\null

\end{document}